\documentclass[aps,twocolumn,prl,showpacs]{revtex4}
\usepackage{graphicx}
\usepackage{amsfonts} 
\usepackage{amssymb,amsmath}

\begin{document}
%\begin{titlepage}
\title{Hubbard ring: currents induced by change of magnetic flux}
 \author{M.  Mierzejewski, J. Dajka and J.  {\L}uczka }
\address{Institute of Physics, University of Silesia, 40-007
Katowice, Poland }

\begin{abstract} 

We investigate   currents in a quantum ring threaded by a  magnetic flux which  can be  varied 
 in an arbitrary way from an initial value $\phi_i$ at time $t_i$ to a final value 
$\phi_f$ at time $t_f$.  Dynamics of electrons 
in the ring is described by the Hubbard and the extended Hubbard models.  
We demonstrate that  time dependence of the induced current 
bears information on  electron correlations. 
In the case of the Hubbard model with infinite on--site repulsion we prove that the current
for $t>t_f$ is independent of  the  flux variation before $t_f$.  Additionally, 
this current is  fully determined by a solution of the initial equilibrium problem and the
value of $\phi_f$. 
Apart from mesoscopic rings our results pose important implications for designing of quantum motors built out as the ring--shaped optical lattice.
\end{abstract}

%}
\maketitle
%%%%%%%%%%%%%%%%%%%%%%%%%%%%%%%%%%%%%%%%%%%%%%%%%%%%%%%%%%%%%%%%%%%%%%%%%%%%
%
Time--dependent manipulation of  quantum states in  nanosystems 
is an important problem  directly related to future applications both in the context of quantum control  \cite{control} and the reduction of decoherence \cite{decohcontrol}. There is a natural ground  for implementing such ideas: mesoscopics, where quantum effects play a crucial role \cite{brandes}. 
Unfortunately, analysis of quantum systems affected by external time--dependent
force or field $F(t)$ is extremely difficult and only very few models are exactly solvable.
The best known examples concern a driven quantum oscillator \cite{husimi,perelomov,zerbe}
and a two--level system in circularly polarized magnetic field \cite{rabi,rev_rab}. In many cases  the solvability is related to certain symmetries of the system. 
Moreover,  there are interesting regimes of strong  external driving  when   
 the linear response theory  cannot be applied. 
For periodically driven systems one may apply the Floquet theory \cite{kohler}, however, 
in many cases this general approach provides only approximate numerical results.
 Another problem concerns the possibility
of experimental verification of theoretical predictions. Here a pertinent question arises:
 how does the final (generally non--equilibrium) state depend 
on a particular shape of $F(t)$? For the purpose of practical applications it
is desirable to find systems which are robust against small temporal
changes of $F(t)$, e. g., originating  from imperfect realization of the assumed conditions.  
  A realistic example will be pointed out in this work. 

We consider  one--dimensional (1D) mesoscopic systems of ring topology  threaded 
by a magnetic flux $\phi$.  Dynamics of electrons moving in the ring is described by the  Hubbard 
and the extended Hubbard models. 
We propose a simple scheme involving  variation of  the  magnetic flux  to manipulate  currents: 
it  is changed  from its initial value $\phi(t)= \phi_i$ for time $ t \le t_i$ to the final value 
$\phi(t)=\phi_f$ for $t \ge t_f$.  
We show that for  vanishingly small and infinitely 
strong many--body interactions the resulting  current 
does not depend on the way the magnetic flux is modified or switched on. In other words,  
the current for $t \ge t_f$ is entirely determined by the solution of the equilibrium problem at $t=t_i$. 
Although the method of reasoning   we apply for non-interacting fermions is rather trivial, it nicely illustrates the general method that  is applicable also to a non--trivial case of
 correlated  electron systems.   
Our analysis can be applied either to  nonsuperconducting 
mesoscopic rings or to  rings built in the optical lattice setup \cite{hanggi_ring}. 
The difference in  energy scales in both  systems  shows up mainly in 
different time scales of the external driving.  

We   start with the Hamiltonian of non--interacting  particles in the  ring threaded by  a magnetic flux. 
It is  a sum of  one-particle Hamiltonians
\begin{eqnarray}\label{ring1}
H(t)=\frac{1}{2m}\left[p - \frac{e}{L}\phi(t)\right]^2,
\end{eqnarray}
where $L$ is the circumference of the ring and $e$ is a charge of the particle.  
 The current operator is related to the momentum observable in the following way 
\begin{eqnarray}
I(t)= -\frac{\partial H(t)}{\partial \phi(t)} =    \frac{e}{mL}\left[p-\frac{e}{L}\phi(t)\right]. 
\end{eqnarray}
Now, let us assume that the magnetic flux $\phi$ is varied   in an arbitrary way from an initial value $\phi_i$ at time $t_i$ to a final value  $\phi_f$ at time $t_f$.  
One can  explicitly extract the time--dependent part of the current operator 
\begin{eqnarray}
I(t)= I_i+\Delta I(t), 
\end{eqnarray}
%
%where
%
\begin{eqnarray}
I_i=\frac{e}{mL}\left[p-\frac{e}{L}\phi(t_i)\right],  \\
\Delta I(t)= \frac{e^2}{mL^2}\left[\phi(t_i)-\phi(t)\right]. 
\end{eqnarray}
The averaged current flowing in the ring   is determined by the relation 
\begin{eqnarray}\label{curr}
\langle I(t) \rangle = \mbox{Tr}\left[\rho(t)I(t)\right], 
\end{eqnarray}
where $\rho(t)$ is a density matrix of the system.  Its  time evolution  is determined by the von Neumann equation $i \hbar \dot{\rho}(t)=[H(t),\rho(t)]$. Note that 
from Eq. (\ref{ring1}) it follows that 
\begin{eqnarray}\label{com}
\left[H(t_1),H(t_2)\right]=0
\end{eqnarray}
 for arbitrary $t_1$ and $t_2$.  In consequence, a solution of the von Neumann equation has the form 
\begin{eqnarray}\label{vonN}
\rho(t) = \mbox{exp}\left[-\frac{i}{\hbar} \int_{t_0}^t H(s) ds\right]  \rho(t_0)  
\mbox{exp}\left[\frac{i}{\hbar} \int_{t_0}^t H(s) ds\right].
\end{eqnarray}
If  $\rho(t_0)$ commutes with the Hamiltonian $H(t)$  for any $t>t_0$  then $\rho(t) = \rho(t_0)$. 
The latter requirement is not very restrictive. It is fulfilled by the Gibbs state 
$\rho(t_0) \propto \exp\left[-\beta H(t_0)\right]$ and, from the experimental point of view, seems to be the easies and the most natural choice of the initial preparation. 
Let us take $t_0=t_i$ and $\rho_i =\rho(t_i)$. Then the averaged current reads 
\begin{eqnarray}\label{obser}
\langle I(t) \rangle = \mbox{Tr}\left[\rho(t)I(t)\right]=\mbox{Tr}[\rho_i I_i]+  \Delta I(t),  
\end{eqnarray}
where the first term on r.h.s. is the  initial equilibrium persistent current and  the second term is 
the  current induced by the  time-dependent component 
$\Delta I(t) $. As the latter quantity is independent of the initial state, one can easily
calculate the current induced in a system of  
$n$ non--interacting particles $\Delta I_n(t)=n \Delta I(t)$. It is instructive to compare
$ \Delta I_n(t) $ with  the amplitude of persistent currents $I_{PC}$ at zero temperature \cite{pers}. 
One finds     
\begin{eqnarray}\label{stosun}
\frac{\langle \Delta I_n(t)\rangle}{I_{PC}}=2\frac{\phi(t_i)-\phi(t)}{\phi_0},
\end{eqnarray}
where $I_{PC}=v_F e/L$, $v_F=\hbar \pi n/(mL)$ is the Fermi velocity  and  $\phi_0=h/e$ is the  flux quantum.
 
Let us notice remarkable properties resulting from Eq.(\ref{stosun}): 
({\em 1}) The averaged current $ \langle I(t) \rangle $ depends on the flux only at the same instant of time $t$, i.e.,
the current is independent of the way how the magnetic flux is switched on; 
({\em 2}) 
$ \langle I(t) \rangle $
is fully determined by
the solution of the initial equilibrium problem,  i.e.  by the equilibrium 
mean value   $\mbox{Tr}[\rho_i I_i]$;
({\em 3}) one can induce currents which are significantly larger in amplitude than the persistent currents provided that 
$\phi(t)-\phi(t_i) \gg \phi_0$. 

It is known that  electrons in 1D systems are almost always strongly 
correlated \cite{nagaosa}. In that sense, the free electron approximation applied to  1D quantum rings is   
disputable. 
In the following we discuss  to what extent the conclusions derived for the model of noninteracting particles  
are applicable to  more realistic systems of correlated particles.  
Here we consider a  1D ring  described by the  'standard model' of correlated electrons, i.e.  
by the  Hubbard model \cite{nagaosa,hubb}, 
\begin{eqnarray}\label{hubbard}
H_{\mathrm H}(t)=-J\sum_{j,\sigma}\left( \mbox{e}^{i\tilde{\phi}(t)} a^\dagger_{j+1,\sigma}a_{j,\sigma}+ h.c.\right)+ U\sum_j n_{j,\uparrow}n_{j,\downarrow}, \nonumber \\
\label{hub}
\end{eqnarray}
where  $J$ is the hopping integral, $U$ is the on--site Coulomb repulsion,  
 $n_{j,\sigma}=a^\dagger_{j,\sigma}a_{j,\sigma}$ and $\sigma=\uparrow,\downarrow$.
For the ring consisting of $N$ sites $\tilde{\phi}(t)=2 \pi \phi(t)/N \phi_0 $. The current operator  reads 
\begin{equation}
I(t)= i \frac{2 \pi J}{N \phi_0} \sum_{j,\sigma} \left(\mbox{e}^{i\tilde{\phi}(t)} a^\dagger_{j+1,\sigma}a_{j,\sigma}- h.c.\right). 
\end{equation}
We  choose $J$ as the energy unit, whereas time and current will be expressed in units of  $\tau=\hbar/J$ and $I_0=2 \pi J/N \phi_0$, respectively. 
% 
%%%%%%%%%%%%%%%%%%%%%%%%%%%%%%%%%%%%%%%%%%%%%%%%%%
\begin{figure}[b]
\begin{center}
\includegraphics[width=0.45\textwidth,angle=0]{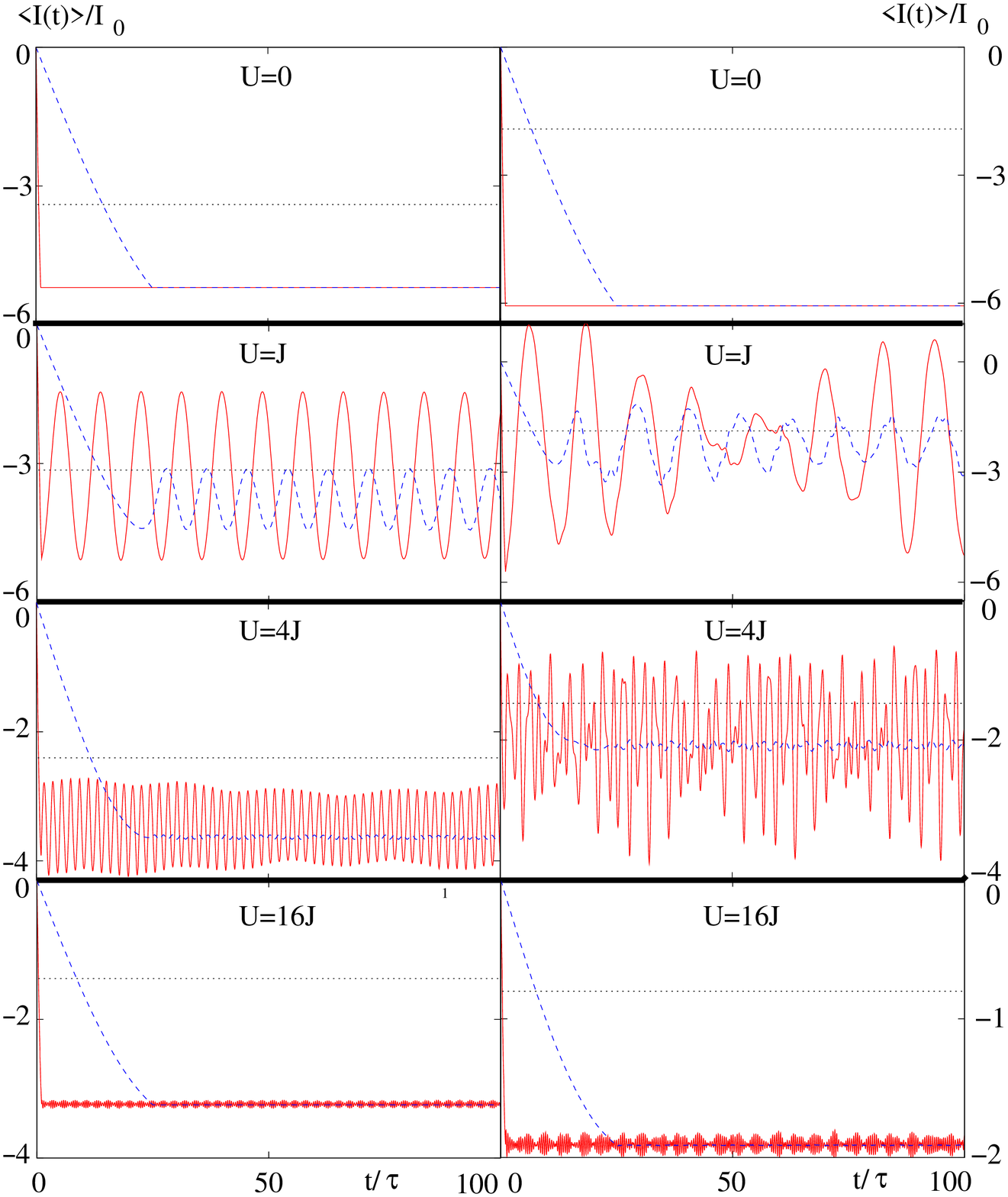}
\end{center}
\caption{(color online) The averaged current $\langle I(t) \rangle$ induced by  change of the  magnetic flux $\phi(t)$ from  $\phi(t_i=0)=0$ to  $\phi(t_f)=\phi_0$ with  the change rate $\dot\phi(t) =$ const for $t\in (t_i, t_f)$. The results are  obtained for $N=6$, $n_{\uparrow}=2$, $n_{\downarrow}=2$ (left panels) and  $N=6$, $n_{\uparrow}=3$, $n_{\downarrow}=2$ (right panels).
Continuous (red) lines show results for $t_f=\tau$ and 
dashed (blue) lines for $t_f=25 \tau$.
The horizontal dotted lines show  
maximal values of the equilibrium persistent current.
}
\label{fig1} 
\end{figure}
%%%%%%%%%%%%%%%%%%%%%%%%%%%%%%%%%%%%%%%%%%%%%%%%%%%

Before we carry out  discussion based on analytical results,  it is instructive to inspect  numerical  
studies.
Fig. 1 illustrates the time-dependence of the average current {$\langle I(t) \rangle$ obtained from 
Eq. (\ref{curr}) by  a solution of the von Neumann equation for $\rho(t)$  under  an  initial condition being the equilibrium state $\rho_i \propto  \exp\left[-\beta H_{\mathrm H}(t_i)\right]$ with $\beta \rightarrow \infty$. The presented results have been obtained  for a ring consisting 
of $N=6$ sites with various numbers of spin up  $n_{\uparrow}$ and spin down $n_{\downarrow}$ particles.
Apart from the case $n_{\uparrow}+n_{\downarrow}=N$ (when the system is insulating) 
the qualitative results are independent of the system size and  the number of particles. 
We find that in the  small and large $U/J$-limits, a dc current is observed for $t>t_f$. Its amplitude is independent of the  rate of flux variation and  is greater than the equilibrium persistent current. For moderate values of $U/J$, the current displays time-oscillations. However, its  average over time  is non-zero and the dc component can be detected.  
Frequency of oscillations depends  on electron correlations:  for  stronger correlations, i.e.,  for larger $U/J$,  frequency of the current is higher. On the contrary, its amplitude decreases as $U/J$ increases. 
 The amplitude of oscillations  is more  sensitive to the rate of the flux changing 
 $\dot\phi= (\phi_f-\phi_i)/(t_f-t_i)$: slow changes of the flux   result in  small  amplitudes of the current oscillations and {\it vice versa}.   The detailed analysis of currents (e.g., regular {\it vs.} chaotic behavior) in this intermediate  regime will be presented elsewhere. 
 
The numerical results suggest that the conclusions formulated for  free particles  [({\em 1})-({\em 2})] hold true also for the system described by 
the Hubbard Hamiltonian (\ref{hub}) with $U=0$ or $U/J=\infty$. The former case ($U=0$) 
is again trivial, since in the Bloch representation one gets
\begin{eqnarray}\label{tight}
H_{\mathrm H}(t)=-\sum_{k,\sigma} 2 J \cos(k-\tilde{\phi}(t))a^\dagger_{k,\sigma}a_{k,\sigma}.
\end{eqnarray}
As Eq.  (\ref{com}) holds true for the above Hamiltonian, 
$\rho(t)=\rho(t_i)={\mathrm const}$ and $\langle I(t) \rangle$ is independent
of the magnetic flux $\phi(t')$ for $t'<t$. In the following we prove that it also holds
true for  $U/J=\infty$. In the case of infinitely strong Coulomb repulsion one can rewrite
the Hamiltonian (\ref{hub}) in the form  
 \begin{eqnarray}\label{hubb1}
H_{\mathrm H}(t)&=& H_L(t)+H_R(t),  \\
H_L(t)&=&H^{\dagger}_R(t)= -J \mbox{e}^{i\tilde{\phi}(t)} P \sum_{j,\sigma} a^\dagger_{j+1,\sigma}a_{j,\sigma} P, 
\end{eqnarray}  
where the operator $P= \Pi_{j=1}^N (1-n_{j,\uparrow}n_{j,\downarrow})$ projects out states with doubly occupied sites. It is clear that
$[H_L(t_1),H_L(t_2)]=[H_R(t_1),H_R(t_2)]=0$. Then, in order to prove that Eq.(\ref{com}) 
holds true, it is enough to show that $[H_L(t_1),H_R(t_2)]=0$.  One finds
\begin{eqnarray}
&& [H_L(t_1),H_R(t_2)]= J^2 \mbox{e}^{i\left(\tilde{\phi}(t_1)-\tilde{\phi}(t_2) \right)}(A-B), \label{kom3} \\
&& A= \sum_{i,j,\sigma,\mu} P a^\dagger_{i+1,\sigma}a_{i,\sigma} \bar{P} 
  a^\dagger_{j,\mu}a_{j+1,\mu}P \label{A1},  \\
&& B= \sum_{i,j,\sigma,\mu} P a^\dagger_{j,\mu}a_{j+1,\mu}  \bar{P}  a^\dagger_{i+1,\sigma}a_{i,\sigma}P. \label{A2}
\end{eqnarray}  
We have introduced the notation  $\bar{P}\equiv P$  to distinguish different positions of these operators.
Analyzing the operator $A$ one can note that 
because of the presence of the projection operators $P$ one can neglect $\bar{P}$ unless 
the hopping $a^\dagger_{i+1,\sigma}a_{i,\sigma}$
 removes double occupancy generated by
$a^\dagger_{j,\mu}a_{j+1,\mu}$. Similar method of reasoning applies to the operator $B$. 
Therefore, in Eqs. (\ref{A1}) and (\ref{A2}) one can
replace $\bar{P}$ with $(1-\delta_{ij})+ \delta_{ij}\bar{P}$.  Note that 
$a_{i,-\sigma} \bar{P}   a^\dagger_{i,\sigma} |\psi \rangle=0 $ for arbitrary
state $|\psi \rangle$. Taking  into account these properties
one gets
\begin{eqnarray}
A-B&=&  \sum_{i,j,\sigma,\mu} (1-\delta_{ij}) P [ a^\dagger_{i+1,\sigma}a_{i,\sigma}, a^\dagger_{j,\mu}a_{j+1,\mu}] P \nonumber \\
&& + \sum_{i \sigma}P a^\dagger_{i+1,\sigma}a_{i,\sigma} \bar{P} 
  a^\dagger_{i,\sigma}a_{i+1,\sigma}P \nonumber \\
&& -  \sum_{i \sigma}P a^\dagger_{i,\sigma}a_{i+1,\sigma}\bar{P} a^\dagger_{i+1,\sigma}a_{i,\sigma}P
\end{eqnarray}
The first term vanishes because the commutator is proportional to 
$\delta_{ij}$. Now, the projection operators $\bar{P}$ in the second and third terms
can be replaced by $1-n_{i,-\sigma}$ and  $1-n_{i+1,-\sigma}$, respectively. Then, one gets
\begin{eqnarray}
&&A-B= P  \sum_{i \sigma} n_{i+1,\sigma}(1-n_{i,\sigma})(1-n_{i,-\sigma})P \nonumber \\
&& - P  \sum_{i \sigma} n_{i,\sigma}(1-n_{i+1,\sigma})(1-n_{i+1,-\sigma})P.
\end{eqnarray} 
One can see that $A-B$ is expressed as a difference of two operators (first and second lines in the above equation). The first one
counts how many occupied sites {\em succeed} empty sites, whereas the latter one counts how
many occupied sites {\em precede} empty sites. In a ring--shape system these numbers
are equal for an  arbitrary state. Hence $A-B=$0, next  $[H_L(t_1),H_R(t_2)]=0$ and then Eq. (\ref{com}) holds true.

It is important to compare our method with the standard approach to 1D Hubbard  model 
via the Bethe Ansatz \cite{wulib}. The latter one provides a solution
for all interaction strengths and band-fillings and we refer to Ref. 
\cite{voit} for a comprehensive review on the
equilibrium properties of the 1D Hubbard model. In particular, 
the elementary excitations are expressed in terms of holons and spinons which, in general  interact and are not independent \cite{voit}. 
Then,  analysis of correlation functions within the Bethe 
Ansatz is by far not straightforward. The complete charge--spin separation 
over all energy scales occurs only in the case  $U \rightarrow \infty$ what significantly simplifies calculations of the correlation functions
 \cite{uinfty}. Although our approach goes beyond the the linear response theory 
and does not relay on the Bethe Ansatz, the results for $U \rightarrow \infty$ can be interpreted  in terms 
of the charge--spin separation. As the vector potential
couples to charged orbital degrees of freedom, in the case of the full spin--charge separations the system
responds to time--dependent flux in the very same way as a system of non--interacting fermions.
Within this conjectural interpretation one comes to a conclusion that observation of 
currents induced by time--dependent fluxes gives important information on the spin-charge separation.
   %%%%%%%%%%%%%%%%%%%%%%%%%%%%%%%%%%%%%%%%%%%%%%%%%%%%%%
\begin{figure}[htpb]
\begin{center}
\includegraphics[width=0.45\textwidth,angle=0]{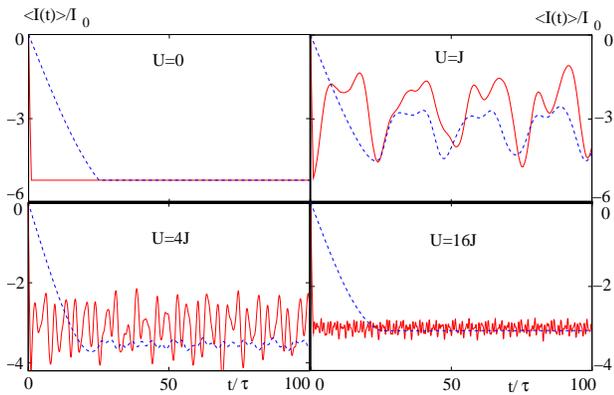}
\end{center}
\caption{The same as in the left column of Fig. \ref{fig1} but for the extended
Hubbard model with $V=0.4U$.
}
\label{fig2} 
\end{figure}
 %%%%%%%%%%%%%%%%%%%%%%%%%%%%%%%%%%%%%%%%5%%%%%       

Finally,  we  verify whether the above discussion can be generalized for a system that cannot
be solved via the Bethe Ansatz. For this sake we have carried out numerical
calculations for the extended Hubbard model with the Hamiltonian \cite{penc}
\begin{equation}
H_{\mathrm{EH}}=H_{\mathrm H} + V\sum_{j\sigma \mu} n_{j,\sigma}n_{j+1,\mu}.  
\end{equation}
     Fig. \ref{fig2} shows similar results to those presented in the left column 
of Fig. \ref{fig1} but calculated for the extended Hubbard model with $V=0.4U$.
We are unable to prove that the general properties of $\langle I(t) \rangle$  derived
for the non--interacting system hold true also for the extended Hubbard model in the limit
of strong interactions. However, a comparison of Figs. \ref{fig1} and \ref{fig2} strongly suggests that
it is actually the case.  Fluctuations of $\langle I(t) \rangle$ for $t>t_f$  gradually extinguish when
the interaction becomes stronger and the magnitude of these oscillations decreases when $t_f$ increases.
Simultaneously, for $U \gg J$ the magnitude of the current for $t > t_f$
becomes almost independent of $t_f$. We have carried out similar calculations for
$V=J$ ($V$ is independent of $U$) and found that the above conclusions concerning the role of $U$
 remain valid.   

In conclusion, we have  analyzed  currents induced by  temporal changes of an external  magnetic  flux piercing a quantum ring.  
On the one hand,  experimental observations of the current may give important insight into various fundamental  properties of the system, as e.g.  electron correlations and the spin--charge separation. On the other hand, by controlling the rate  of  change of  the external magnetic flux,  we demonstrate how  an oscillating current of   desired amplitude and frequency can be induced. 
Moreover, its    time-average is non-zero and contains a dc component.  The significant advantage of the 
method based on the magnetic flux variation is the 'noninvasive' manipulation performed outside the ring, without coupling to external leads. 
%Experimental verification of our results in mesoscopic rings might be difficult 
Recent progress  in the highly controlled fabrication of quantum ring 
structures makes the verification  of our findings  quite realistic in  the  nearest future. 
We should note that basic  limitations concerning  mesoscopic systems (time shorter than the relaxation time and the flux of  order of the flux quantum)  are not so restrictive for experiments performed in the optical lattice setup. 
In this context, our results pose some important implications for the design of quantum motors discussed in Ref. 
\cite{hanggi_ring}. For $U/J \ll 1$ or $U/J \gg 1 $  significant currents can be generated
neither by impulses with $\phi(t_f)\simeq \phi(t_i)$ nor by a magnetic flux 
that has small time--averaged value $\overline{\phi(t)}\ll \phi_0$. Consequently, in the case of 
the ac--driven quantum motors  (considered, e.g., in Ref. \cite{hanggi_ring}) the systems with $U \sim J$ should provide
the best performance.  Generation of  significant currents 
in a system with $U/J \ll 1$ or $U/J \gg 1 $ is possible provided $\phi(t_f)- \phi(t_i) \sim \phi_0$.
In the latter case, magnitude of the current is independent of the way how the magnetic flux 
is modified. This feature may  facilitate the experimental realization.  Summarizing,  our  proposal  
expose  new prospects of inducing,  controlling and  manipulating  of   currents in nonsuperconducting  
quantum small systems of ring topology.  

%\section*{Acknowledgment} 
The work  supported by  
the Polish Ministry of Science and Higher Education 
under the grant N 202 131 32/3786.


\begin{thebibliography}{50}
\bibitem{control} J. Werschnik and E.K.U. Gross, J. Phys. B: At. Mol. Opt. Phys. {\bf 40}, 175 (2007). 
\bibitem{decohcontrol} G. Gordon,  J. Phys. B: At. Mol. Opt. Phys. {\bf 42}, 223001 (2009). 
\bibitem{brandes} T. Brandes, Phys. Rep. {\bf 408}, 315(2005). 
\bibitem{husimi} K. Husimi, Progr. Theor. Phys. {\bf 9}, 381 (1953).
\bibitem{perelomov} A. M. Perelomov  and V. S. Popov, 
Theor. Mat. Fiz. {\bf 1}, 275 (1970) [Sov. Phys. JETP, {\bf 30}, 910 (1970)].
\bibitem{zerbe} C. Zerbe and P. H{\"a}nggi, Phys. Rev. E {\bf 52}, 1533 (1995).
\bibitem{rabi} I. I. Rabi, Phys. Rev. {\bf 51}, 652 (1937).
\bibitem{rev_rab} P. K. Aravind and I. O. Hirshfeld, J. Phys. Chem. {\bf 88},
4788 (1984). 
\bibitem{kohler} S. Kohler {\it et al}, Phys. Rep. {\bf 406}, 379 (2005).
P. H\"anggi in {\it 	
Quantum transport and dissipation} (Wiley-VCH, Weinheim, 1998). 
  \bibitem{hanggi_ring} A. V. Ponomarev, S. Denisov, and P. H\"anggi, Phys. Rev. Lett. {\bf 102}, 203601 (2009).
\bibitem{pers} I. O. Kulik, JETP Lett. {\bf 11}, 275 (1970); M. B\"{u}ttiker, Y. Imry, and R. Landauer, Phys. Lett. A {\bf 96}, 365 (1993); 
P. Mohanty, Ann. Phys. {\bf 8} 549 (1999); U. Eckern and P. Schwab, J. Low Temp. {\bf 126} 1291 (2002).  
\bibitem{nagaosa} N. Nagaosa {\it Quantum field theory in strongly correlated electronic systems} (Springer, Berlin, 1999).
%\bibitem{stare} Stare prace o PC: Landauer,Buttiker, Imry, Gefen i ska. 
%\bibitem{imam} Y. Yamamoto and A. Imamoglu, {\it Mesoscopic quantum optics}, Johm Willey, New York (1999). 
\bibitem{hubb} Fabian H.L. Essler, Holger Frahm, Frank Göhmann, Andreas Klümper and Vladimir E. Korepin, {\it The One-Dimensional Hubbard Model} (Cambridge University Press, Cambridge, 2005).
%\bibitem{nonpers} V. E. Kravtsov and B. L. Altshuler, Phys. Rev. Lett. {\bf 84}, 3394 (2000).
%\bibitem{pers} I. O. Kulik, JETP Lett. {\bf 11}, 275 (1970); M. B\"{u}ttiker, Y. Imry, R. Landauer, Phys. Lett. A {\bf 96}, 365 (1993); P. Mohanty, Ann. Phys. {\bf 8} 549 (1999); U. Eckern and P. Schwab, J. Low Temp. {\bf 126} 1291 (2002) 
%\bibitem{hanggi} Hanngi: quantum transport and dissipation
%\bibitem{kohler} S. Kohler {\it et al}, Phys. Rep. {\bf 406}, 379 (2005).
%P. H\"anggi in {\it 	Quantum transport and dissipation} (Wiley-VCH, Weinheim 1998). 
%\bibitem{adiab} przybizenie adiabatyczne i recent controvereses
\bibitem{wulib} E. H. Lieb and F. Y. Wu, Phys. Rev. Lett. {\bf 20}, 1445 (1968).
\bibitem{voit} J. Voit,  Rep. Prog. Phys. {\bf 58}, 977 (1995).
\bibitem{uinfty} M. Ogata and H. Shiba, Phys. Rev. B {\bf 41}, 2326 (1990); 
B. Kumar Phys. Rev. B {\bf 70}, 155121 (2009).
\bibitem{penc} For $U/J \to\infty$,  the extended Hubbard model can be mapped onto the model 
of sipnless fermions with nearest-neighbour repulsion which, in turn,  can be mapped onto the XXZ Heisenberg 
model and then  analyzed via the Bethe Ansatz. For details, see e.g., 
F. D. M. Haldane Phys. Rev. Lett. {\bf 45}, 1358 (1980);  F. Mila and K. Penc, J. Electron Spectrosc.  
{\bf 117-118}, 451 (2001).
  
\end{thebibliography}
\end{document}